\documentclass{aa}
\usepackage{amssymb}
\usepackage{graphicx}
\usepackage{hyperref}

\begin{document}

\def\psr{PSR~J0030+0451}

%\sloppy

%  ------ To make boxes wider ---------
\def\wideru{ \vrule height 2.8ex   width 0ex depth 0ex}
\def\widerul{\vrule height 2.5ex width 0ex depth 0ex}
\def\widerd{ \vrule height 0ex   width 0ex depth 3ex}

\def\twolines#1#2{
\renewcommand{\arraystretch}{0.8}
\begin{tabular}{@{}c@{}}
#1 \vrule height3.2ex width0ex \\ #2 \\[.7ex]
\end{tabular}
}

\title{Deep BVR Imaging of the Field of the Millisecond Pulsar PSR~J0030+0451
with the VLT\thanks{Based on observations performed at the European Southern 
Observatory, Paranal, Chile (ESO Programme 67.D-0519).}}
\author{A.B.~Koptsevich\inst{1} 
\and P.~Lundqvist\inst{2}
\and N.I.~Serafimovich\inst{1,2}
\and Yu.A.~Shibanov\inst{1}
\and J.~Sollerman\inst{2}
%\and G.G.~Pavlov\inst{3}
}
\institute{
Ioffe Physical Technical Institute, Politekhnicheskaya 26,
St. Petersburg, 194021, Russia
\and 
Stockholm Observatory, AlbaNova, Department of Astronomy,
SE-106 91 Stockholm, Sweden
}
\date{Received --- Oct 30, 2002, accepted --- Dec 12, 2002} 

\authorrunning{A.~Koptsevich et al.}
\titlerunning{Deep BVR imaging of the field of PSR~J0030+0451}

\offprints{A.B.~Koptsevich;\hfill\\
e-mail: kopts@astro.ioffe.rssi.ru}

\abstract{
We report on deep BVR-imaging of the field of the nearby millisecond
pulsar PSR~J0030+0451 obtained with the ESO/VLT/FORS2. 
We do not detect any optical counterpart down 
to $B \gtrsim 27.3$, $V \gtrsim 27.0$ and $R \gtrsim 27.0$ 
in the immediate vicinity of the
radio pulsar position. The closest detected sources are offset 
by $\gtrsim 3\arcsec$, 
and they are 
excluded as counterpart candidates by our astrometry.
Using our upper limits in the optical, and including 
recent {\sl XMM-Newton\/} X-ray data we show 
that any nonthermal power-law spectral  
component of neutron star magnetospheric origin, as 
 suggested by the interpretation of X-ray data, must be suppressed
 by at least 
a factor of $\sim 500$ 
in the optical range. 
This either rules out the nonthermal 
interpretation or suggests a dramatic spectral 
break in the  $0.003-0.1$~keV range of the power-law spectrum.
Such a situation  has never been 
observed  
in the optical/X-ray spectral region
of ordinary pulsars, and the origin of 
such a
break
is unclear. 
An alternative interpretation with a purely thermal 
X-ray spectrum  is consistent with our optical upper limits.
In this case the X-ray emission is dominated 
by  hot polar caps of the pulsar. 
\keywords{pulsars: general -- pulsars, individual: PSR~J0030+0451  --  
stars: neutron}
}
\maketitle

\section{Introduction}

Millisecond pulsars (hereafter MSPs) differ from ordinary  
radio pulsars by much shorter spin periods $P$, smaller period   
derivatives $\dot{P}$, 
higher dynamical ages $\tau$, weaker magnetic 
fields $B$, and 
evolution histories (see, e.g., recent review by \cite{Lor01}).
Contrary to ordinary pulsars, 
only 9 of 56 MSPs currently known in the Galactic disk
and 25 of 52 MSPs found in globular clusters 
are isolated objects (\cite{Lor01}; \cite{Lor02}); \cite{Pos01}).
It is believed that the fast rotation of these neutron stars (NSs) 
was 
gained
in the past by angular momentum transfer 
during mass accretion from 
a companion star (\cite{BH91}).
This was supported by the discoveries of three accretion-powered 
X-ray MSPs in low-mass X-ray binaries (e.g., SAX J1808.4-3658, see
\cite{WvdK98}).
   
\begin{table*}[t]
\caption{Parameters of \psr\ (Lommen et al.\ \cite{L00} and \cite{Beck00},
unless specified otherwise).}
\label{t:param}
\begin{tabular}{ccccccccccc}
\hline\hline
\multicolumn{6}{c}{Observed}&&\multicolumn{4}{c}{Derived} \widerul\\
\cline{1-6}\cline{8-11}
$P$ &
$\dot P$ &
$\alpha_{2000}$, $\delta_{2000}^{a,b}$ & 
$\mu_\alpha$, $\mu_\delta^b$ &
$l$, $b^d$ &
$D\!M^e$ &&
$\tau$ &
$B$ &
$\dot E$ & 
$d^f$ \wideru \\
ms &
$10^{-20}$ &&
mas yr$^{-1}$ &&
cm$^{-3}$ pc &&
Gyr & 
G  &
erg $s^{-1}$ &
pc \widerul \\ 
\hline 
4.865 &
$1.0\pm0.2$ &
\twolines{$00^h30^m27$\fs432(9)$^c$}{04\degr51\arcmin39\farcs67(2)} &
\twolines{$4.1\pm3.4$}{$-24.6\pm8.0$} &
\twolines{113\fdg1}{$-$57\fdg6} &
$4.3328(2)$ &&
7.7 &
$2.2 \times 10^{8}$ &
$3.4 \times 10^{33}$ &
230 \widerul\\
\hline
\end{tabular}
$^a$~Coordinates are at the epoch of the VLT observations, MJD 52134 (Aug 13, 2001) 
\ \ \ \ \ $^d$~Galactic coordinates \wideru\\
$^b$~Updated values of the proper motion (A.~Lommen 2001, private communications)  
\ \ \ \ \ \ $\!$ $^e$~Dispersion measure \\
$^c$~Numbers in parentheses are uncertainties referring to the last significant digit quoted \\
$^f$~$D\!M$-based distance (the new distance model by \cite{cl02} places the pulsar to 317 pc) \\ 
\end{table*} 

Despite these differences, the distribution of integrated radio 
luminosities, as well as the luminosity dependence on $P$, $\dot{P}$, $B$, 
and spindown energy losses $\dot{E}$, are apparently similar for these 
much older and low-magnetized NSs, and for ordinary pulsars 
(\cite{KL01}).
About a dozen radio MSPs
have been detected in X-rays. It is remarkable that their efficiency 
in converting spindown energy to X-ray luminosity is roughly the same
as for ordinary pulsars, $L_{\rm X}/\dot{E}\sim 10^{-3}$ 
(\cite{BT97}; \cite{Beck00}).
This suggests  that the emission mechanisms responsible 
for the multi-wavelength radiation of MSPs and  ordinary 
pulsars  can be
similar, and one could therefore expect to detect MSPs
in other spectral ranges as well, as has been done for several 
ordinary pulsars. Detection of the first MSP
in gamma-rays (\cite{K00}) supports this idea.
   
To our knowledge, there are still no reports on optical detection of
isolated MSPs.
It is hardly possible 
to detect thermal emission from the entire surface of these 
old, $10^{8}-10^{10}$ yr, and cold NSs.  However, the spindown energy, 
expected to power the nonthermal emission of pulsars, can be 
much higher for MSPs than for old ordinary pulsars, 
and may even rival that of  young pulsars. Assuming the same efficiency 
of conversion of spindown energy to nonthermal optical luminosity as for 
ordinary pulsars, one can estimate that nearby MSPs may well 
be detectable in the optical with large telescopes. A problem is, however, 
that most of the nearby and energetic MSPs are components 
of close binary systems where the companion is predominantly either  a white 
dwarf or main sequence star (\cite{Lor01})  
which outshines the pulsar in the optical. 
Fortunately, there are at least nine {\em solitary} MSPs
in Galactic disk
whose companions are believed to have been either 
evaporated or ablated (see, e.g., Lommen et al.\ \cite{L00}). 

Here we report on deep BVR imaging of the field of one of these 
solitary millisecond pulsars, PSR J0030+0451.
This pulsar was only recently discovered with the Arecibo telescope 
(Lommen et al.\ \cite{L00}), and soon thereafter
detected in X-rays during the final observations 
with the {\sl ROSAT/PSPC}  (\cite{Beck00}). 
Recently it was re-observed in X-rays 
with the {\sl XMM-Newton} (\cite{BA02}).  
This relatively nearby NS 
(see Table~\ref{t:param} for its parameters)
is characterized by
high X-ray flux, about $(4.3-6.8)\times10^{-13}$ erg s$^{-1}$ cm$^{-2}$ in 
the $0.1-2.4$~keV band, and low interstellar absorption,
$N_{\rm H}\la 3\times10^{20} $ cm$^{-2}$, 
corresponding to a color excess 
$E(B-V)\la 0.06$ mag. 
This makes it a 
promising candidate for 
optical detection. 
In Sect.~2 we present the observations and the data reduction.
In Sect.~3 we discuss our 
results in the optical in conjunction with the available X-ray data.

\section{Observations and data reduction}

\label{s:corr}

The field of \psr\ was observed in service mode on July 26 and August 13, 2001,
with the FOcal Reducer/low dispersion Spectrograph (FORS2) on the ESO/VLT/UT2
telescope\footnote{See http://www.eso.org/instruments/fors/ for details on
the instrument.}, with a pixel scale of 0\farcs2. We used Bessel filters for
B and V, and an ESO special filter for R (R$_{\rm special}$, henceforth called
R$_{\rm s}$)\footnote{See~http://www.eso.org/instruments/fors1/Filters/.}.
Unfortunately, some of the V images were corrupted by bad CCD columns near the 
expected position of the pulsar and were not used in the analysis. 
The images we used are listed in Table~\ref{t:obslog}.

\begin{figure*}[t]
\setlength{\unitlength}{1mm}
\begin{picture}(178,84)(0,0)
\put (  0, 0)   {\includegraphics[width=89mm,bb=105 220 470 560,clip]{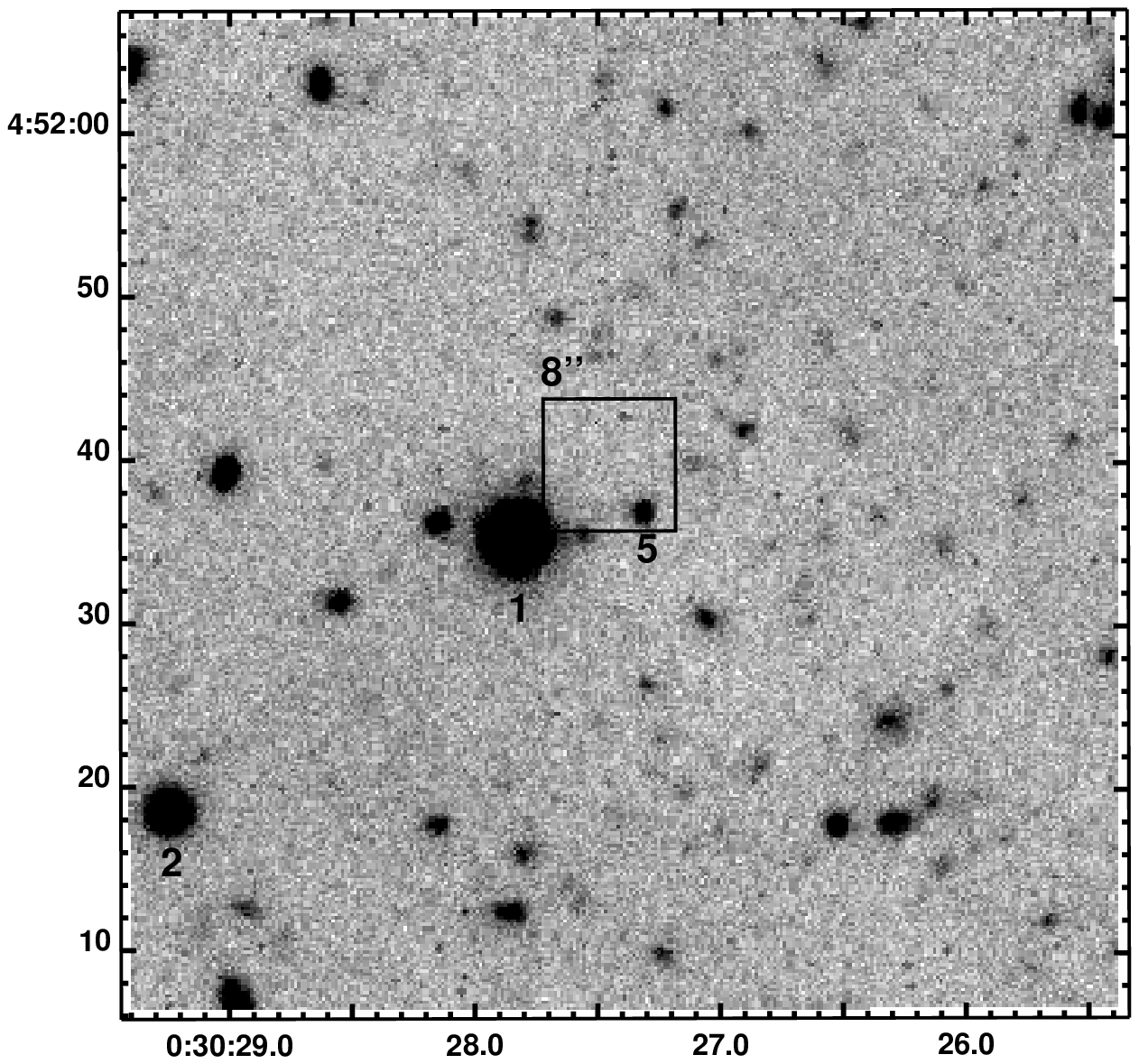}}
\put ( 90,53)   {\includegraphics[height=29mm,bb=160 260 440 540,clip]{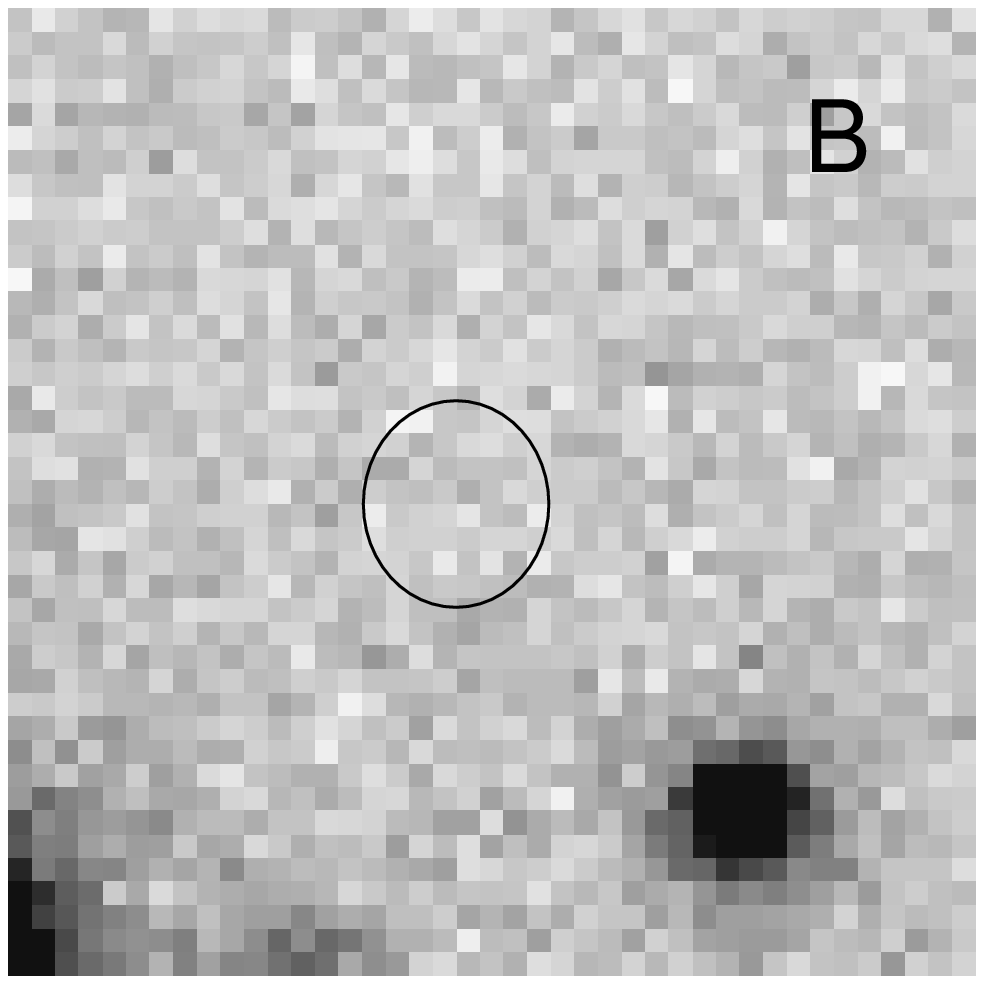}}
\put (120,53)   {\includegraphics[height=29mm,bb=160 260 440 540,clip]{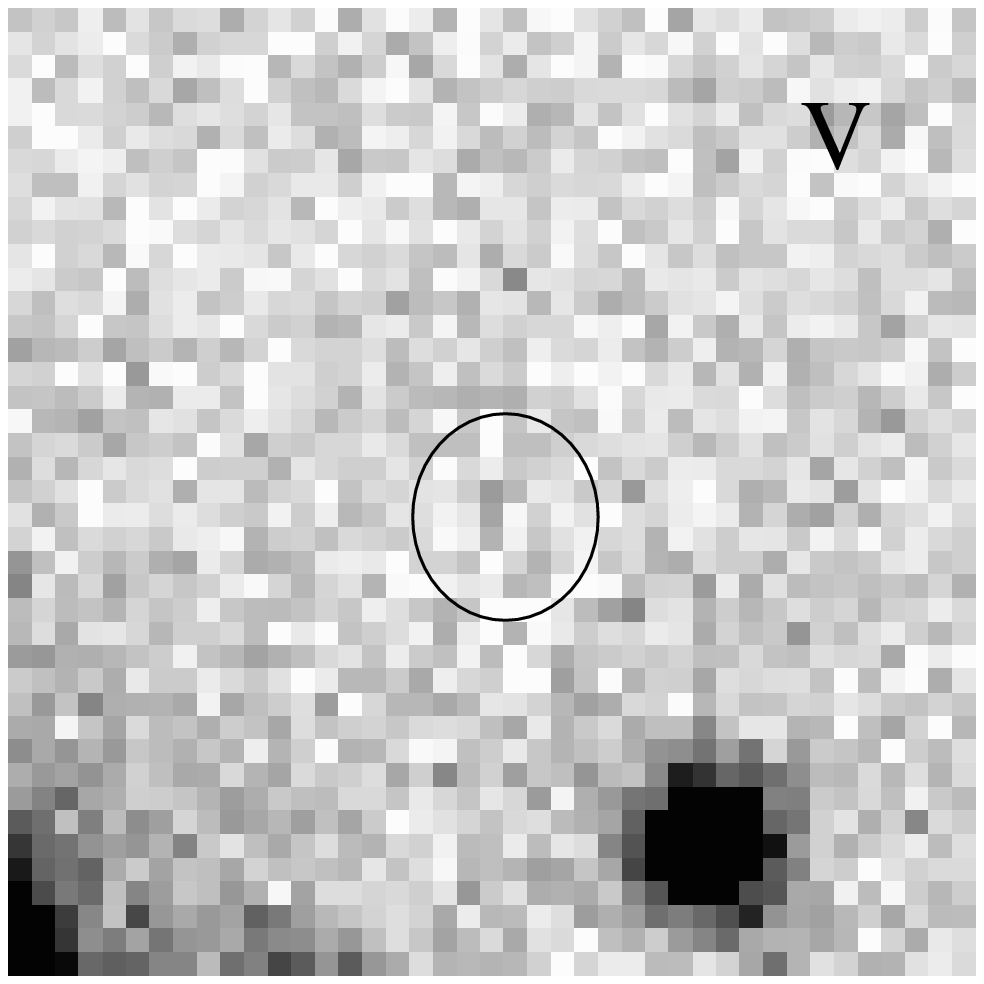}}
\put (150,53)   {\includegraphics[height=29mm,bb=160 260 440 540,clip]{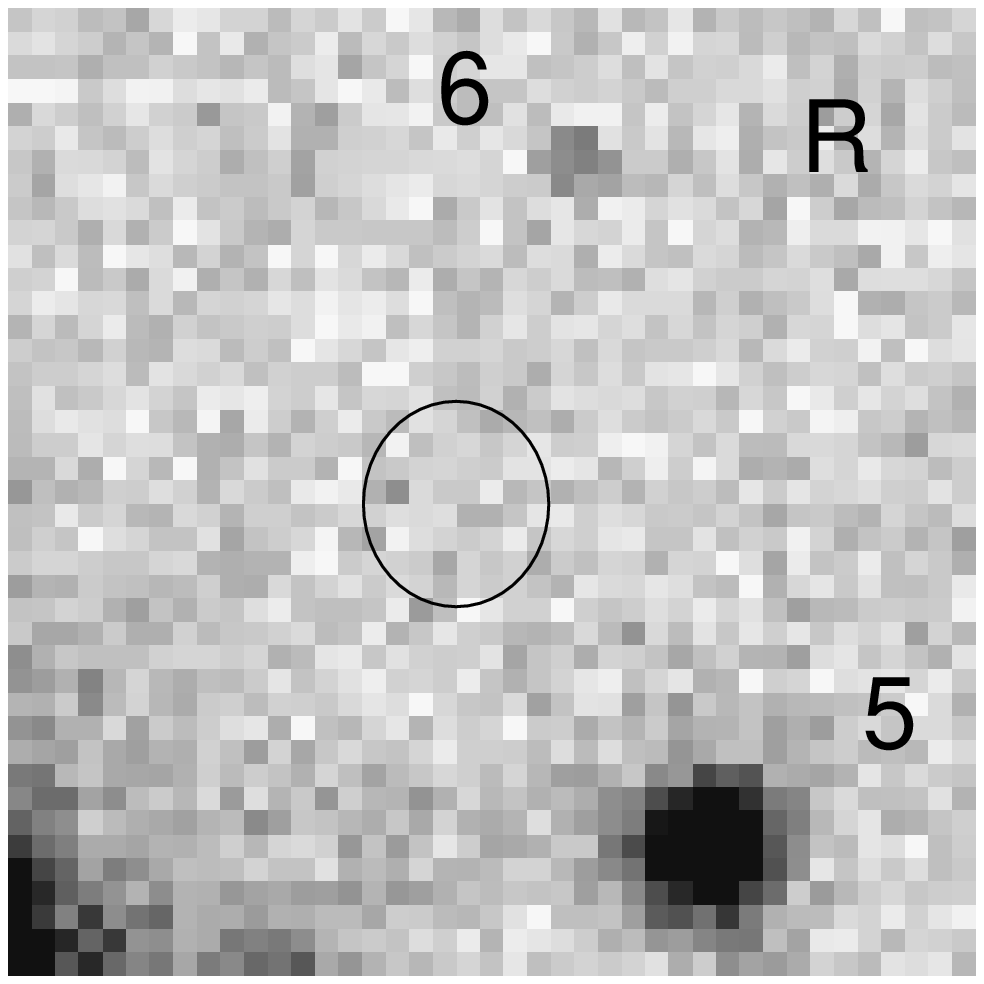}}
\put ( 93, 1)   {\parbox[b]{85mm}{\caption{%
{\bfseries\itshape Left panel:} 
$1\arcmin\times1\arcmin$ overview of
the \psr\ field extracted from the $6\farcm8\times6\farcm8$ frame of the VLT/FORS2 
image in the R$_{\rm s}$ band. The central box is blown-up in the 
{\rm smaller panels\/}.
A nearby bright USNO star (object~1) and two other stellar 
objects (2 and 5 from Table~\ref{t:mesh}) are marked. 
{\bfseries\itshape Smaller panels:} 
$8\arcsec\times8\arcsec$ pulsar vicinity in the BVR$_{\rm s}$ bands.
$3\sigma$ uncertainty ellipses of the expected 
pulsar position at the epoch of the VLT observations 
are marked. No pulsar is detected in these frames. Two objects are marked 
``5'' (referring to Table~\ref{t:mesh}) and ``6'' in the R$_{\rm s}$ image, and are 
discussed in the text.    
} \label{f:ima1} }}
\end{picture}
\end{figure*}

Bias subtraction and flatfielding were performed in a standard way,
and the reduced individual images were aligned using a set of bright, 
non-saturated field stars. Standard utilities from the NOAO {\sf IRAF} package
were then used to combine the images applying the averaged sigma clipping 
algorithm {\sf avsigclip} with the {\sf scale} parameter equal to {\sf none}.
The pulsar vicinity is shown for B, V and R$_{\rm s}$ bands 
in Fig.~\ref{f:ima1}.

\begin{table}[t]
\caption{Log of VLT observations of \psr\ in the BVR$_{\rm s}$ bands.
Exposure time is 720 s for all frames.} 
\label{t:obslog}
\begin{tabular}{lccccc}
\hline
No  & Date     & Band & Time  & Airmass & Seeing  \\ 
    & UT       &      & UT    &         & arcsec  \\ 
\hline 
1   & 26.07.01 & B    & 09:40 & 1.172   & 0.6 \\
2   &          &      & 09:53 & 1.188   & 0.5 \\
\cline{2-6}
3   & 13.08.01 & B    & 04:55 & 1.579   & 1.1 \\ 
4   &	     &      & 05:09 & 1.495   & 1.1 \\
5   &	     &      & 06:34 & 1.208   & 0.8 \\
6   &          &      & 07:56 & 1.150   & 0.7 \\
7   &	     &      & 08:09 & 1.155   & 0.9 \\
\cline{3-6}
8   &	     & V    & 05:38 & 1.358   & 0.7 \\ 
9   &	     &      & 05:52 & 1.311   & 0.8 \\
10  &	     &      & 06:06 & 1.269   & 0.8 \\
11  &	     &      & 07:14 & 1.159   & 0.7 \\
12  &	     &      & 07:28 & 1.151   & 0.6 \\
13  &	     &      & 07:42 & 1.149   & 0.7 \\
\cline{3-6}
14  &	     & R$_{\rm s}$& 06:47 & 1.187   & 0.7 \\
15  &	     &      & 08:24 & 1.167   & 0.7 \\
16  &	     &      & 08:37 & 1.181   & 0.7 \\
17  &	     &      & 08:53 & 1.205   & 0.7 \\
18  &	     &      & 09:07 & 1.230   & 0.7 \\
\hline
\end{tabular}
\end{table}

\begin{table*}[t]
\caption{Magnitudes of several stars in the \psr\ field 
with coordinates and  offsets from the pulsar position 
shown in the table. 
Stars 1, 2 and 5 are marked within the left frame in Fig.~\ref{f:ima1}, while 3 and 4 are outside this frame.}
\label{t:mesh}
{\footnotesize
\begin{tabular}{ccccccc}
\hline\hline
No & $\alpha_{2000}$ & $\delta_{2000}$ & Offset & $B$ & $V$ & $R_{\rm s}$ \\
\hline
1 & 00:30:27.83 & +04:51:35.2  & $\ \ \ \ $7\farcs5 & 19.20$\pm$0.01 & $-$ & $-$ \\
2 & 00:30:29.25 & +04:51:18.3  & $\ \ \,$34\farcs6 & 22.32$\pm$0.02 & 20.92$\pm$0.01 & 20.27$\pm$0.01 \\
3 & 00:30:39.33 & +04:53:15.4  & 3\arcmin21\farcs9 & 23.77$\pm$0.06 & 22.08$\pm$0.01 & 21.29$\pm$0.01 \\
4 & 00:30:25.39 & +04:53:14.2  & 1\arcmin39\farcs3 & 24.15$\pm$0.07 & 22.40$\pm$0.02 & 21.78$\pm$0.01 \\
5 & 00:30:27.32 & +04:51:36.9  & $\ \ \ \;$3\farcs2 & 23.82$\pm$0.05 & 23.69$\pm$0.06 & 23.63$\pm$0.06 \\
PSR & 00:30:27.43 & +04:51:39.67 & $\ \ \ \ \ \ \,-$ & $\ge$27.3      & $\ge$27.0      &$\ge$27.0 \\  
\hline
\end{tabular}\\
}
\end{table*} 

The radio position of \psr\ at the epoch of the VLT observations  
(for which 13.08.01 was adopted, see Table~\ref{t:param})
was determined using recent radio ephemerides (A.~Lommen 2001, 
private communication for additional Arecibo observations). 
Astrometrical referencing of our images was made with {\sf IRAF} 
tasks {\sf ccmap/cctran} using the positions of several dozens of  
reference stars from the USNO A-2.0 catalogue seen 
in the images and optimizing the astrometrical  fit 
by removing step by step the reference stars with the largest  residuals.   
For the 5 most suitable stars we finally 
got 1$\sigma$ rms-errors of 0\farcs05 and 0\farcs11, 
and maximum residuals of 0\farcs20 and 0\farcs24, in RA and Dec, respectively.  
Combining the rms-errors with the nominal USNO accuracy of 0\farcs24  and 
radio ephemeris uncertainties, we obtained the 3$\sigma$ pulsar VLT/FORS 
position uncertainties 0\farcs79 and 0\farcs88 in RA and Dec, respectively.
The resulting 3$\sigma$ error ellipse is marked in the smaller  images in 
Fig.~\ref{f:ima1}. No reliable counterpart candidate 
was detected within, or close to the error ellipse 
of the expected pulsar position. 
 We also double-checked the astrometry with 5 stars from the GSC-II catalog.
The obtained rms errors were 0\farcs05 and 0\farcs11 in RA and Dec, respectively,
and the expected pulsar position was moved 0\farcs06 west and 0\farcs14 north 
in respect to the USNO position. 
However, since the main source of errors  is the catalog uncertainty,
we accept the USNO results as more conservative estimate.

For the photometric calibrations we used the photometric standards 
from the PG1323$-$085 and SA109 fields (\cite{Landolt}), observed 
at the second night of our observations, 
and average Paranal extinction coefficients\footnote{
http://www.eso.org/observing/dfo/quality/FORS2/qc/ 
photcoeff/photcoeffs\_fors2.html}.
We then derived $3\sigma$ detection limits as
$m=-2.5\log\left(3\sigma\sqrt{A}/T_{\rm exp}\right)+m_0$, where
$\sigma$ is the standard deviation of the flux in counts per pixel,
$T_{\rm exp}$ is the exposure time, $A$ 
is the area of an aperture (in pix$^2$) with a radius of 
1$\arcsec$ (corresponding to $\sim 83$\% of the flux in a PSF of our images),  
and $m_0$ is the photometric 
zeropoint, including corrections for atmospheric extinction.  
The limits are: $B = 27.3$, $V = 27.0$, and $R_{\rm s} = 27.0$.   
 
In Table~\ref{t:mesh} we list BVR$_{\rm s}$ magnitudes for several of the 
objects in the \psr\ field.
Object 1 is the USNO star U0900\_00118426 marked in 
the left panel of Fig.~\ref{f:ima1}. 
It is oversaturated 
in our V and R$_{\rm s}$ images. Object 5 is 
the source seen at the bottom right of the blown-up images 
in Fig.~\ref{f:ima1}.
This is the closest object to the radio position of the pulsar    
clearly detected in all 
our images. 
Although it is too blue to be a normal star,
it cannot be considered a pulsar counterpart candidate because of the large 
offset (3\farcs4) from the radio position, 
provided the radio astrometry is as accurate as claimed by Lommen et al.\ (\cite{L00}). 
Like most other blue objects     
in the field at this magnitude level (we found at least five objects roughly 
of the same color), it is most likely extragalactic, and 
similar to  blue objects found in the HDF images. The object marked  ``6'' in the 
  upper right of the R$_{\rm s}$
 image in Fig.~\ref{f:ima1}, is only 
 marginally detected at $R_{\rm s}$=27.45$\pm$0.61. It is not detected in the other bands
 and may well be an artefact of the reductions. It is also
too far (3\farcs1) from the radio pulsar position to be an 
optical counterpart candidate.

\section{Discussion}
\label{s:disc}

\begin{figure*}[t]
\includegraphics[width=130mm, angle=-90, clip]{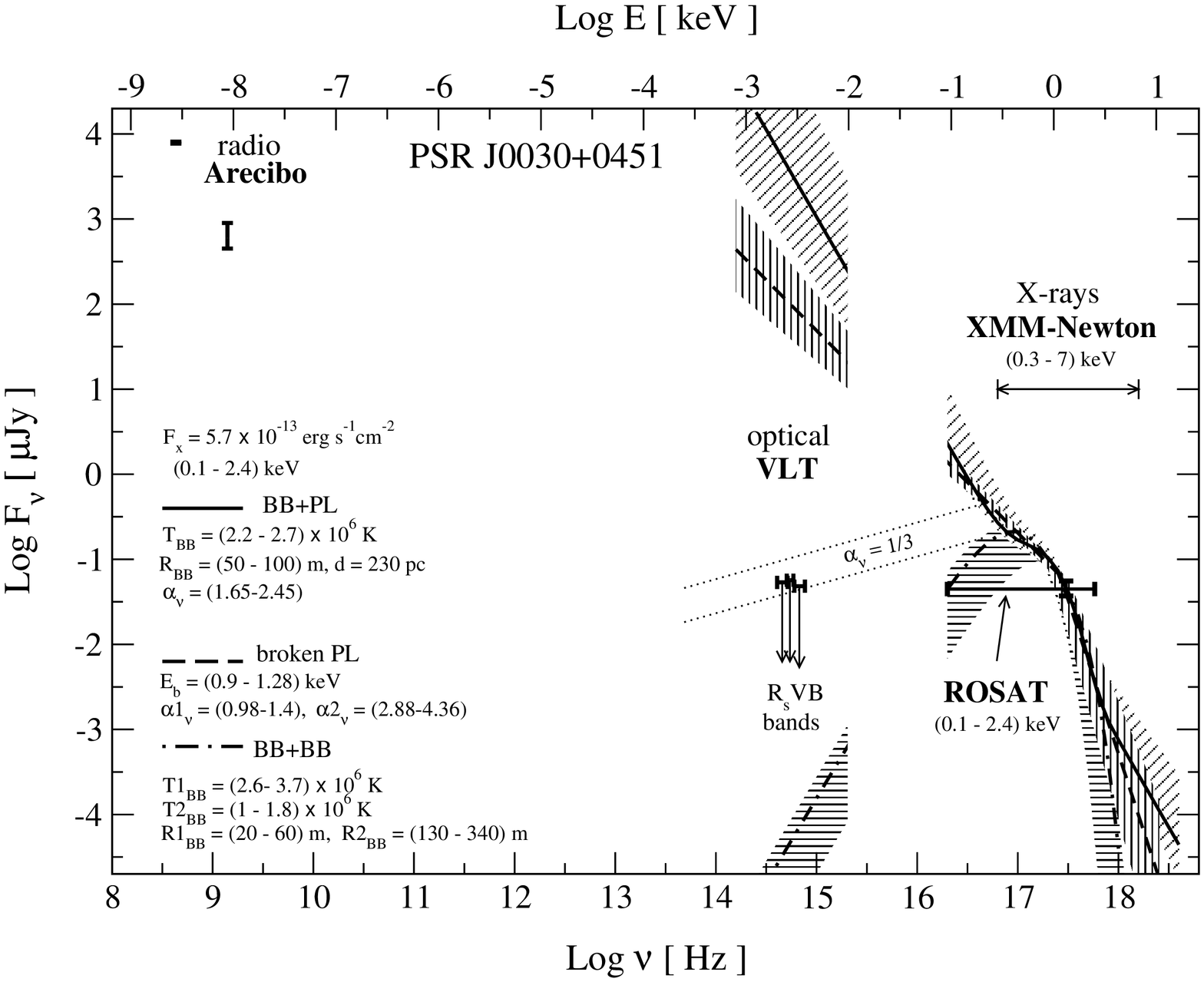}
\caption{
Radio and X-ray observations of \psr\ with the Arecibo telescope     
(Lommen et al.\ \cite{L00}), {\sl ROSAT} (\cite{Beck00}), 
and {\sl XMM-Newton} (\cite{BA02}), 
 as well as VLT upper limits in the BVR${\rm_s}$ optical bands. 
Solid, dashed, and dot-dashed lines show the best spectral fits 
of the  {\sl XMM-Newton} data with three alternative 
two component  spectral models: 
BB+PL,  
broken PL,  
and 
BB+BB,  
respectively (BB stands for blackbody).  
Parameters of the fits and a mean unabsorbed integrated flux 
$F_{\rm X}$, approximately the same for all the fits    
(\cite{BA02}), are indicated in 
the plot. All fits are acceptable at the same 
significance level in the $\sim$ $0.3-7$~keV range, indicated by a horizontal bar 
with arrows at the upper-left,    
and  their $\sim$90\% uncertainties are shown as shaded regions.  
The unabsorbed model spectra are extrapolated toward the optical range.
Previous {\sl ROSAT} data contain no spectral information and 
provide only an integrated flux indicated in the figure by a 
bold cross (\cite{Beck00}), 
roughly consistent with the more recent {\sl XMM-Newton} results.         
 The VLT upper limits, marked by end bars, 
 show  that  
any nonthermal PL  component obtained from 
the above spectral fits of the X-ray data must be strongly suppressed 
in the optical range. This implies  either a cutoff or a strong 
break 
in the $0.003-0.1$~keV range. The purely thermal BB+BB model is consistent 
with our optical upper limits. Further details of this plot are discussed in Sect.~\ref{s:disc}. 
} \label{f:flux}
\end{figure*}

 Based on the apparent similarity  of the radiation properties  
 of MSPs and ordinary pulsars, and assuming 
 the same mechanisms to generate optical emission 
 in both types of pulsar,
 we expected to detect the optical counterpart of \psr\ 
 at the $m\sim26$ visual magnitude level, assuming
 a simple scaling of $\dot{E}/d^2$ from known optical fluxes 
 of ordinary pulsars.
 Our observations were deeper, but still we
 did not detect any reliable counterpart candidate.
           
To try to understand what our optical non-detection implies,
we have 
plotted the available information about the 
multiwavelength spectrum of \psr\ including radio, optical, and X-ray data in Fig.~\ref{f:flux}. 
For the X-ray region we have included preliminary results of recent 
{\sl XMM-Newton} observations (\cite{BA02}) where 
the pulsar was clearly detected in the $0.3-7$~keV range. The 
detection range is shown by 
a horizontal bar with arrows in Fig.~\ref{f:flux}.
The overall spectrum compares well with multiwavelength spectra of 
ordinary pulsars (e.g., \cite{Kop01}), i.e., the pulsar 
flux is higher in the radio and fades toward the X-ray range. 
This can 
be explained by different emission mechanisms in the radio (coherent) and
at shorter wavelengths (non-coherent). However, a more detailed inspection
of the optical/X-ray range reveals a feature which 
has not been seen for ordinary pulsars. For the latter, the optical flux 
is usually close to an extrapolation of the nonthermal high energy tail 
of the  X-ray emission, usually described by a power-law (PL). From Fig.~\ref{f:flux}  
it is clear that this is not the case for the \psr. 

Fig.~\ref{f:flux} shows that in the {\sl XMM-Newton}  range  
the data can be fitted equally well by three different two-component spectral models: 
a blackbody + power-law model (BB+PL), a broken (or curved) PL model, or a 
model based on two different blackbodies (BB+BB) with $N_{\rm H} \le 2.5\times10^{20}$ cm$^{-2}$. 
From the X-ray data alone it is difficult to discriminate    
between the three models, although the sharp X-ray pulse profile perhaps 
favors the domination of a nonthermal PL component (\cite{BA02}). 
The VLT upper limits allow additional constraints using an 
extrapolation of the X-ray model spectra toward the optical range.

From Fig.~\ref{f:flux} it is seen that the low-energy extension  
of the BB+PL fit 
overshoots 
the optical flux upper limits of \psr\ by $\sim$5 orders 
of magnitude. The broken PL extension, which implies 
a flatter spectrum in the soft X-ray energy band,  
is also 3-4 orders of magnitude higher. 
We have
not corrected the data for
interstellar extinction, which is low and 
does not play any significant role at such large 
differences. 
The strong suppression of the PL components 
in the optical range suggests  that these two models either  
should be ruled out, or that their PL components 
must  have a strong 
break or even a cutoff at a photon energy  
somewhere in the $0.003-0.1$~keV range. 
Such a situation has never been observed 
for any 
ordinary pulsar. 
For example, the spectral index $\alpha_{\nu}$ (defined 
as $F_{\nu} \propto \nu^{-\alpha_{\nu}}$)
of the young Crab pulsar changes from $\sim0.5$ for soft X-rays 
to zero in the FUV/optical/near-IR range 
(e.g., \cite{S00}; \cite{S02}). For the relatively young  
Vela-pulsar (\cite{MC01}), the
middle-aged PSR B0656+14 (\cite{Kop01}) and PSR B1055$-$52 
(\cite{P02}), and even for the old ordinary pulsars 
PSR B0950+08 (\cite{Zhar02}) and PSR B1929+10 (\cite{M02}), 
the extension of the PL X-ray component matches  
the optical flux, suggesting the same mechanism 
of nonthermal emission in the optical and X-ray ranges.  
It is not clear what could be the reason for such a strong change in the 
nonthermal spectral slope of \psr\ from negative in X-rays to positive in 
the optical range. 
 As seen from Fig.~\ref{f:flux}, the X-ray PL fits and our upper limits 
exclude any flat extension toward the optical range,  
even if we place a break point 
energy 
at the lower boundary of the  
{\sl XMM-Newton} range $\sim 0.1$~keV.
Note that the low-frequency extensions of the PL components 
overshoot even the radio fluxes. 
This is also not typical for 
ordinary pulsars. 
One can  assume that nonthermal emission 
is  due to  synchrotron  radiation  
of relativistic particles in the magnetosphere 
of the pulsar.  
In this case we obtain for the simplest  
monochromatic particle distribution over the 
energy a  spectral flux of $F_{\nu} \propto \nu^{1/3}$
in the low frequency range below a maximum frequency 
$\nu_{\rm m} \propto B\gamma^2$. Here 
$B$ is the magnetic field, and $\gamma$ is the gamma-factor 
of the emitting particles. From the spectral shape suggested by the 
X-ray data and optical upper limits 
it is natural to put $\nu_{\rm m}$ near the maximum of the X-ray 
spectral flux at the low boundary of the {\sl XMM-Newton} range.
At typical gamma-factors of primary and secondary relativistic particles 
in pulsar magnetospheres, i.e. $\sim 10^6$ and $\sim 10$, respectively, 
and  for the period of \psr\ $\sim4.9$~ms,   
the same peak frequency value is predicted   
by a model of synchrotron emission from the pulsar 
light cylinder suggested  by \cite{Mal01}.
For the expected synchrotron flux  
below the adopted $\nu_{\rm m}$ value
(see Fig.~\ref{f:flux}, dotted lines) we 
would likely hint the optical counterpart, 
but we did not.

The purely thermal BB+BB spectral model  
is consistent with our upper limits without any additional assumptions.
Its Rayleigh-Jeans tail is about 6 stellar magnitudes fainter in the 
optical than our upper limits, and would hardly be detectable with 
present telescopes. 
 Thermal photons can be emitted by hot polar 
caps of the pulsar. The two-blackbody fit  indicates a non-uniformity in 
the temperature distribution over the caps. 
This could be described by
heat propagation over the surface of 
the neutron star out of a hot cap core, including  
neutron star atmosphere effects, as has been done in the case 
of the MSP J0437$-$4715 (\cite{Z01}).
An additional faint PL component is required 
to fit an excess over the thermal emission at high energy  X-rays
from PSR J0437$-$4715.  If the same would be true for \psr, it 
could be brighter in the optical  than estimated from the simple BB+BB model
due to a contribution from a similar nonthermal component of magnetospheric origin. 
Deeper X-ray observations are probably needed to detect 
this component  in the high energy tail of the  \psr\ spectrum. 
The similarity of X-ray and radio pulse 
profiles of this pulsar suggests that radio and X-ray peaks 
are in phase (\cite{BA02}),  
although direct timing to confirm this 
has not yet been done. In the frame work of the thermal model this means 
that radio emission is generated close to the polar cap surface 
and the similarity of the pulse shapes is likely caused by the same 
geometry of the emitting regions.

Based on our upper limits we can constrain also   
the efficiency of converting spindown power of  \psr\  
to optical emission.    
The luminosity in the B band is 
$L_{\rm B}=4\pi d^2F_{\rm B} \Delta\nu_{\rm B} \la 6.3 \times 10^{26}~d_{230}^2$ erg s$^{-1}$, 
and hence the optical efficiency 
$\eta_{\rm B} =L_{\rm B} /\dot{E} \la 1.9\times 10^{-7} d_{230}^2$. 
Here $d_{230}=d/230$~pc is the normalized distance to \psr.
This upper limit is about half the efficiency of the middle-aged 
PSR~B0656+14 (at 500 pc) and exceeds the efficiencies of the Geminga and 
Vela pulsars by about 1 and 2 orders of magnitude, respectively 
(see, e.g., \cite{Zhar02}). It is interesting to note that  
the expected efficiency in the BB+BB model in Fig.~\ref{f:flux} is 2-3 orders of 
magnitude lower than the upper limit derived above from our optical data, 
and that the efficiency in the BB+BB model is comparable to that of the Vela 
pulsar.  The comparison of the efficiencies makes sense  
here, since the thermal emission from hot polar caps 
and the nonthermal magnetospheric emission of the Vela-pulsar, 
both are powered by relativistic particles produced in magnetospheres 
of rapidly rotating NSs.  Thus, we see that the optical efficiency 
of the \psr, as derived from our and X-ray observations, 
is not unusually low, but is compatible with the
efficiency range of ordinary pulsars detected in the optical band. 
 
It has been shown for ordinary pulsars that spectral index 
appears to became steeper with pulsar age in the optical range 
(\cite{Kop01}; \cite{MC01}), while it flattens 
in gamma rays (e.g., \cite{sg02}).  
It has been noticed also across a restricted set of young and middle-aged 
pulsars detected in the optical and gamma regions that the gamma-ray 
efficiency increases with age, while the reverse is true 
for the optical efficiency (\cite{gold95}).
This would suggest that there is a reprocessing of the
gamma-photons into the optical
in pulsar magnetospheres, and that it is more efficient for
younger pulsars than for
older ones (e.g., \cite{sg02}). Thus, it would be not 
surprising that very old \psr\ is fainter in the optical 
than we expected.     
However, recent optical studies of old ordinary pulsars  
(\cite{Zhar02}; \cite{M02}; \cite{M02b})
have revealed nonmonotonous behavior of the optical efficiency {\it vs} age  
with a minimum at  $\tau \sim 10^4-10^5$~yr and further increase 
towards higher ages $\tau \ga 10^7$~yr. Old pulsars can be actually 
much more efficient than the middle-aged ones  
and produce the optical photons with almost the same efficiency 
as young and energetic Crab-like pulsars. 
In this context low optical brightness of the MSP J0030$+$0451
remains puzzling.

A clue to what dominates the X-ray and optical spectrum 
would hopefully be found from observations of \psr\ and 
other MSPs in the FUV and, in particular, in the EUV range. 
Even deep upper limits in these ranges would help to understand how 
strongly the multiwavelength emission and radiation properties of 
MSPs differ from those of ordinary pulsars.

\acknowledgements{We are grateful to Andrea Lommen for access to the yet
unpublished revised radio data on the proper motion of \psr\ and
for useful comments, and to George Pavlov for discussions.
We are also grateful to the anonymous referee for comments
which improved the paper presentation.
Partial support for this work was provided by grant 1.2.6.4 of
the Program ``Astronomia'', and by RFBR (grants 02-02-17668 and 00-07-90183).
Support was also given by The Royal Swedish Academy of Sciences, and
the research of PL is further sponsored by the Swedish Research Council. 
ABK and YuAS are thankful to Stockholm Observatory and The Royal Swedish 
Academy of Sciences for hospitality. ABK also appreciates hospitality of the 
Astronomy Departments of the University of Washington and the Penn State.
PL is a Research Fellow at the Royal Swedish Academy supported by a grant 
from the Wallenberg Foundation. NIS is supported by The Swedish Institute.}

\end{document}